\documentclass[preprint,aps,showkeys,a4paper,showpacs]{revtex4}
\usepackage{epsfig}
\begin{document}
\title{Ensemble equivalence for non-Boltzmannian distributions}
\author{Ra\'ul Toral}
\email{raul@imedea.uib.es} 
\affiliation{Instituto Mediterr\'aneo
de Estudios Avanzados (IMEDEA), CSIC-Universitat de les Illes
Balears\\ Ed. Mateu Orfila, Campus UIB, 07122 Palma de Mallorca, Spain}
\homepage{http://www.imedea.uib.es}

\date{\today}

\begin{abstract} 

We discuss the possibility of using generalized canonical distributions, i.e. using other factors than $\exp(-\beta E)$, in order to compute the equilibrium properties of physical systems. It will be show that some other choices can, in certain cases, lead to a simpler calculation of those properties. The corresponding equivalence between the canonical and the generalized canonical distributions is derived using well-known principles of Statistical Mechanics and we exemplify the method by deriving in a simple way the equilibrium properties of the long--range Ising model.

\end{abstract}

\maketitle
\section{Introduction}

Ensemble equivalence is a key issue in the foundations of Statistical Mechanics. Beyond the physical interpretation, the use of one or another ensemble can simplify the mathematical treatment of a problem. For instance, the Ising model in the absence of a magnetic field can be solved in two dimensions in the canonical ensemble, but not in the microcanonical, and one of the major arguments used in the textbooks to introduce the grand canonical ensemble is that of computing the grand partition function of non interacting quantum gases and derive from it the thermodynamic potentials\cite{balescu}.

As explained in the textbooks, the equivalence between ensembles lies on that fact the probability distributions become effectively, in the thermodynamic limit, a delta function of its arguments. For instance, the canonical energy distribution $P(E)=\Omega(E)\frac{{\rm e}^{-\beta E}}{{\cal Z}}$ (being $\Omega(E)$ the degeneracy of the energy level $E$ and ${\cal Z}$ the canonical partition function) tends to a delta function $\delta(E-U)$ in the thermodynamic limit $N\to\infty$. This is done by checking that the fluctuations
\begin{equation}
\label{sigma}
\sigma^2[E]=\langle E^2\rangle-\langle E\rangle^2
\end{equation}
are negligible in the sense that $\sigma[E]/\langle E\rangle\to 0$ in that limit. A recent review about the necessary mathematical conditions (and some of its violations) for the equivalence between the canonical and the microcanonical ensemble is that of reference \cite{cett05}.

Once the origin of the equivalence between ensembles is understood, it is clear that it should be possible to define many more ensembles that possess the same thermodynamic limit that, say, the microcanonical ensemble. In this paper, we offer a simple way of deriving a family of ensembles which are equivalent to the microcanonical ensemble under some general conditions and are a generalization of the canonical ensemble. We use Jaynes principle of maximization of the entropy as a way of deriving the ensembles. We show how the generalized ensembles can be used in order to derive the thermodynamic potentials. Beyond the ``pedagogical" interest of this result which shows that the Boltzmann distribution can be replaced by others and still obtain the same results for the thermodynamic potentials, there could be cases in which a different ensemble can help to simplify the mathematical treatment of a problem. In this sense, and as a simple application, we rederive the thermodynamic properties of the long range Ising model using an ad-hoc ensemble.

The paper is organized in two main sections: in the next section \ref{section2}, we derive the basic formalism of the generalized ensembles and show the equivalence with the microcanonical ensemble. In section \ref{section2} we compute the thermodynamic potentials of the long-range Ising model.

\section{Derivation of the ensemble}
\label{section1}

Let us consider a system ($N$ particles, volume $V$) with microstates $|m\rangle$ of energy $E_m$. We use Jaynes principle of maximization of the entropy functional\cite{jaynes}
\begin{equation}
\label{spm}
S=-k\sum_m p_m\ln p_m
\end{equation}
to assign probabilities $p_m$ to each of the microstates; $k$ is Boltzmann's constant. Besides the conditions $\sum_m p_m=1$, $p_m\ge 0$ we choose the constrain
\begin{equation}
\label{uphi}
\sum_m p_m \Phi(E_m)=\Phi(U)
\end{equation}
where $\Phi(E)$ is an arbitrary function of the internal energy $U$.

The solution of the maximization problem is a generalized canonical distribution\cite{jaynes} \begin{equation}
p_m=\frac{{\rm e}^{-\beta \Phi(E_m)}}{{\cal Z}}
\end{equation}
where we have defined the {\sl partition function} 
\begin{equation}
{\cal Z}(\beta)=\sum_m{\rm e}^{-\beta \Phi(E_m)}
\end{equation}
In these equations, $\beta$ is a Lagrange multiplier obtained from Eq. (\ref{uphi}):
\begin{equation}
\label{pub}
\Phi(U)=\frac{\sum_m \Phi(E_m) {\rm e}^{-\beta \Phi(E_m)}}{\sum_m{\rm e}^{-\beta \Phi(E_m)}}
\end{equation}
or the equivalent one
\begin{equation}
\label{pub2}
\Phi(U)=-\left(\frac{\partial \ln {\cal Z}}{\partial \beta}\right)_{N,V}.
\end{equation}
The entropy is now computed from Eq. (\ref{spm}) which is equivalent to:
\begin{equation}
\label{szpu}
S=k\left[\ln {\cal Z}+\beta \Phi(U)\right]
\end{equation}
Finally, we need the relation of the parameter $\beta$ with the temperature. By taking the derivative of (\ref{szpu}) with respect to $U$ and noticing that $\beta$ is also a function of $U$ through Eq. (\ref{pub}), we get
\begin{equation}
\left(\frac{\partial S}{\partial U}\right)_{N,V}=k\beta \Phi'(U)
\end{equation}
Which, combined with the thermodynamic relation
\begin{equation}
\label{tsu}
\frac{1}{T}=\left(\frac{\partial S}{\partial U}\right)_{N,V}
\end{equation}
leads us to the identification
\begin{equation}
\label{bfp}
\beta =\frac{1}{kT\Phi'(U)}
\end{equation}
This equation allows to express the potentials $S$ y $U$ in terms of the usual variables $T,N,V$.

It can be proven that this generalized canonical ensemble will lead to the same results than the microcanonical ensemble under general conditions\cite{pathria} which allow to approximate the energy distribution $P(E)=\Omega(E)\frac{{\rm e}^{-\beta \Phi(E)}}{{\cal Z}}$ by a delta function $\delta(E-U)$ in the thermodynamic limit $N\to\infty$. This is done by checking that the fluctuations
\begin{equation}
\label{sigmaphi}
\sigma^2[\Phi]=\langle\Phi^2\rangle-\langle\Phi\rangle^2=-\frac{\partial \Phi(U)}{\partial \beta}
\end{equation}
are negligible in the sense that $\sigma[\Phi]/\langle\Phi\rangle\to 0$ in that limit. A simple algebra leads to an Einstein-like formula:
\begin{equation}
\sigma^2[\Phi]=kT^2C_V\frac{\Phi'(U)^2}{1+TC_V\frac{\Phi''(U)}{\Phi'(U)}}
\end{equation}
relating the fluctuations in the energy with the specific heat at constant volume \makebox{$C_V=\left(\frac{\partial U}{\partial T}\right)_{N,V}$}. Obviously, this and other formulas in this section, reduce to the standard ones if $\Phi(E)=E$, the canonical ensemble.

Another argument in support of the equivalence of the generalized canonical ensemble with the canonical ensemble uses the fact that the probability density function for the energy
\begin{equation}
p(E)=\sum_{E_m=E}p_m=\frac{\Omega(E){\rm e}^{-\beta \Phi(E)}}{\cal Z}
\end{equation}
is, for $\Phi(E)$ an increasing function of the energy, the product of the  rapidly increasing function $\Omega(E)$ and the rapidly decreasing function ${\rm e}^{-\beta \Phi(E)}$. Therefore, $p(E)$ has a very steep peak at some value $E=U^*$. The location of the maximum $U^*$ comes from the equation $\left.\frac{\partial p(E)}{\partial E}\right|_{E=U^*}$ or :
\begin{equation}
\frac{\partial \ln \Omega(U^*)}{\partial U^*}-\beta \Phi'(U^*)=0
\end{equation}
or, using Eq.(\ref{bfp})
\begin{equation}
\frac{1}{kT}=\frac{\partial \ln \Omega(U^*)}{\partial U^*}
\end{equation}
This equation is exactly the same one that, in the microcanonical ensemble, defines the temperature as a function of the internal energy, hence the function $U^*(T,N,V)$ coincides with the internal energy potential $U(T,N,V)$of the microcanonical ensemble.

It is easy to show that the entropy also agrees with that of the microcanonical ensemble. Computing the partition function in Eq.(\ref{szpu}) as a sum over energy values, we obtain:
\begin{equation}
S=k\left[\ln \sum_{E}\Omega(E){\rm e}^{-\beta \Phi(E)}+\beta \Phi(U)\right] 
\end{equation}
We now argue that in the thermodynamic limit the sum will be dominated by its largest term corresponding to $E=U^*$, hence
\begin{equation}
S\approx k\left[\ln [\Omega(U^*){\rm e}^{-\beta \Phi(U^*)}]+\beta \Phi(U^*)\right]=k\ln \Omega(U^*) 
\end{equation}
again in agreement with the microcanonical ensemble.

A simple example will help to check the equivalence. Let us consider the ideal gas in the Boltzmann approximation\cite{balescu}. The sums over states of an arbitrary function $f(E)$ can replaced by integrals
\begin{equation}
\sum_m f(E_m)=\int dE\,f(E)g(E)
\end{equation}
using the energy density
\begin{equation}
g(E)=\frac{[2\pi mh^{-2}V^{2/3}]^{3N/2}}{\Gamma(3N/2)N!}E^{\frac{3}{2}N-1}
\end{equation}

Let us use our generalized ensemble with $\Phi(E)=E^q$ for arbitrary $q>0$. The partition function is:
\begin{equation}
{\cal Z}=\sum_m{\rm e}^{-\beta E^q}=\int_0^{\infty} dE g(E) {\rm e}^{-\beta E}=C(N,V)\beta^{-\frac{3N}{2q}}
\end{equation}
being $C(N,V)$ a function of $N$ and $V$ only. From Eq.(\ref{pub2}) we obtain the relation between $\beta$ and the internal energy $U$:
\begin{equation}
\label{ubeta}
U^q=\frac{3N}{2q\beta}
\end{equation}
The relation between the parameter $\beta$ and the temperature comes from Eq.(\{ref{bpf}):
\begin{equation}
\beta=\frac{1}{kTqU^{q-1}}
\end{equation}
which, combined with Eq.(\ref{ubeta}) leads to
\begin{equation}
U=\frac{3}{2}NkT
\end{equation}
in complete agreement with the microcanonical and canonical results. In this case it is also possible to compute the relative fluctuations as:
\begin{equation}
\frac{\sigma[E^q]}{U^q}=\sqrt{\frac{2q}{3N}}
\end{equation}
which, effectively, tend to zero in the thermodynamic limit $N\to\infty$. 

\section{The thermodynamics of the long-range Ising model}
\label{section2}

The $S_i=\pm 1$, $i=1,\dots,N$ Ising variables interact with a Hamiltonian 
\begin{equation}
{\cal H}=-\frac{2J}{N}\sum_{1\le i<j\le N}S_iS_j
\end{equation}
or
\begin{equation}
{\cal H}=J-\frac{J}{N}\left(\sum_{i=1}^NS_i\right)^2
\end{equation}
The equilibrium properties have been obtained computing the canonical partition function with the help of the Hubbard--Stratonovich transformation and a saddle-point integral\cite{stanley}. We re-derive them by using the generalized canonical ensemble with the choice $\Phi(E)=-(J-E)^{1/2}$. Notice that this is an increasing function for the energy $E\in(J-NJ,J)$. This leads to a partition function 
\begin{equation}
{\cal Z}=\sum_{S_1=\pm 1,\dots, S_N=\pm 1}{\rm e}^{\beta\sqrt{\frac{J}{N}}\sum_{i=1}^NS_i}
\end{equation}
Since the sums are now un-coupled, they are performed trivially:
\begin{equation}
{\cal Z}=\left[2\cosh\left(\beta\sqrt{\frac{J}{N}}\right)\right]^N
\end{equation}
The internal energy follows from (\ref{pub2})
\begin{equation}
\label{intu}
(J-U)^{1/2}=N\sqrt{\frac{J}{N}}\tanh\left(\beta\sqrt{\frac{J}{N}}\right)
\end{equation}
And the relation between the Lagrange multiplier $\beta$ and the temperature $T$ follows from (\ref{bfp})
\begin{equation}
\beta=\frac{2(J-U)^{1/2}}{kT}
\end{equation}
Which, replaced in (\ref{intu}) gives the transcendent equation:
\begin{equation}
(J-U)^{1/2}=\sqrt{JN}\tanh\left(\frac{2}{kT}\sqrt{\frac{J}{N}(J-U)}\right)
\end{equation}
Which is the known result for the internal energy one can obtain by other methods. Defining, $u=\frac{(J-U)}{JN}$, or $u=\frac{-U}{JN}$ in the limit $N\to \infty$, and $T_c=2J/k$:
\begin{equation}
u^{1/2}=\tanh\left(\frac{T_c}{T}u^{1/2}\right)
\end{equation}
from where one gets that non-null values for the energy appear only when $T<T_c$, the critical temperature of the model. Finally, using (\ref{szpu}) we obtain the entropy:
\begin{equation}
\frac{S}{Nk}=\ln\left[2\cosh\left(\frac{T_c}{T}u^{1/2}\right)\right]-\frac{T_c}{T}u
\end{equation}

It is possible to include a magnetic field term $-H\sum_{i=1}^N S_i$ in the Hamiltonian. The function $\Phi(E)$ is chosen such that $\Phi(E)=\sum_{i=1}^N S_i$ and the corresponding sums in the partition function can be easily evaluated. The result can be summarized as follows: Let $x$ be the solution of 
\begin{equation}
h+x=\tanh\left(\frac{T_c}{T}x\right)
\end{equation}
where $h=H/2J$, then the normalized internal energy is
\begin{equation}
u=x^2-h^2
\end{equation}
and the entropy:
\begin{equation}
\frac{S}{Nk}=\ln\left[2\cosh\left(\frac{T_c}{T}x\right)\right]-\frac{T_c}{T}x(x+h).
\end{equation}

In summary, we have introduced a method that uses the equivalence between
ensembles in order to derive the thermodynamic potentials of a Hamiltonian
system using factors other than the Boltzmann factor $\exp{-\beta E}$. We have
applied the method to rederive the equilibrium properties of the long-range
Ising model by using a generalized canonical distribution. It is clear that the
method is of general validity and could be used to derive the equilibrium
properties of other models of Statistical Mechanics, as well as a tool in
numerical simulations of the Montecarlo type.

{\bf Acknowledgments}

This work is supported by the Ministerio de Ciencia y Tecnolog{\'\i}a (Spain)
and FEDER, projects FIS2004-5073-C04-03 and FIS2004-953.

\end{document}